\documentclass[aps,prl, twocolumn,floatfix,nobibnotes]{revtex4-1}
\usepackage{amsmath,amsfonts,amssymb,bbold}
\usepackage{graphicx}
\usepackage{epstopdf}
\usepackage[english]{babel}
\usepackage{natbib}
\usepackage{hyperref}
\usepackage[utf8]{inputenc}

\usepackage{letltxmacro}

\LetLtxMacro{\ORIGselectlanguage}{\selectlanguage}
\makeatletter
\DeclareRobustCommand{\selectlanguage}[1]{%
  \@ifundefined{alias@\string#1}
    {\ORIGselectlanguage{#1}}
    {\begingroup\edef\x{\endgroup
       \noexpand\ORIGselectlanguage{\@nameuse{alias@#1}}}\x}%
}
\newcommand{\definelanguagealias}[2]{%
  \@namedef{alias@#1}{#2}%
}
\makeatother

\definelanguagealias{en}{english}
\definelanguagealias{eng}{english}
\definelanguagealias{English}{english}

\newcommand{\<}{\langle}
\renewcommand{\>}{\rangle}
\providecommand{\abs}[1]{\left\lvert#1\right\rvert}
\providecommand{\norm}[1]{\lVert#1\rVert}

\providecommand{\tr}{{\rm tr}}
\renewcommand{\phi}{\varphi}

\begin{document}
\title{Entanglement and Wigner function negativity of multimode non-Gaussian states}

\author{Mattia Walschaers} 
\email{mattia.walschaers@lkb.upmc.fr}
\affiliation{Laboratoire Kastler Brossel, UPMC-Sorbonne Universit\'es, ENS-PSL Research University, Coll\`ege de France, CNRS; 4 place Jussieu, F-75252 Paris, France}
\author{Claude Fabre}
\affiliation{Laboratoire Kastler Brossel, UPMC-Sorbonne Universit\'es, ENS-PSL Research University, Coll\`ege de France, CNRS; 4 place Jussieu, F-75252 Paris, France}
\author{Valentina Parigi}
\affiliation{Laboratoire Kastler Brossel, UPMC-Sorbonne Universit\'es, ENS-PSL Research University, Coll\`ege de France, CNRS; 4 place Jussieu, F-75252 Paris, France}
\author{Nicolas Treps}
\affiliation{Laboratoire Kastler Brossel, UPMC-Sorbonne Universit\'es, ENS-PSL Research University, Coll\`ege de France, CNRS; 4 place Jussieu, F-75252 Paris, France}

\date{\today}

\begin{abstract}
Non-Gaussian operations are essential to exploit the quantum advantages in optical continuous variable quantum information protocols. We focus on mode-selective photon addition and subtraction as experimentally promising processes to create multimode non-Gaussian states. Our approach is based on correlation functions, as is common in quantum statistical mechanics and condensed matter physics, mixed with quantum optics tools. We formulate an analytical expression of the Wigner function after subtraction or addition of a single photon, for arbitrarily many modes. It is used to demonstrate entanglement properties specific to non-Gaussian states, and also leads to a practical and elegant condition for Wigner function negativity. Finally, we analyse the potential of photon addition and subtraction for an experimentally generated multimode Gaussian state.
\end{abstract}

\maketitle

\paragraph{Introduction ---}Even though the first commercial implementations of genuine quantum technologies are lurking around the corner \cite{QuantumClock,PhysRevLett.117.203003,popkin_quest_2016,sun_quantum_2016,valivarthi_quantum_2016}, much remains uncertain about the optimal platform for implementing quantum functions \cite{obrien_photonic_2009, veldhorst_two-qubit_2015,mohseni_commercialize_2017}. However, it is clear that optics will play a major role in real-world implementations of these technologies \cite{obrien_photonic_2009}. Optical setups have the major advantage \cite{obrien_optical_2007} of being highly robust against decoherence, while also manifesting high clock rates. 

In an all-optical setting, there are various approaches to quantum information protocols, grouped in two classes according to the way information is encoded. Setups which use a few photons, and therefore also rely on single-photon detection to finally extract information, are referred to as discrete variable (DV) approaches. On the other hand, the continuous variable (CV) regime \cite{braunstein_quantum_2005} resorts to the quadratures of the electromagnetic field, ultimately requiring a homodyne detection scheme \cite{armstrong_programmable_2012}. The major advantage of the latter is the deterministic generation of quantum resources, e.g.~entanglement between up to millions of modes \cite{yoshikawa_invited_2016}. Such multimode entangled states, however, remain Gaussian, which implies that their CV properties can be simulated using classical computational resources \cite{bartlett_efficient_2002, rahimi-keshari_sufficient_2016}. Hence, if a quantum information protocol is to manifest a quantum advantage, it requires non-Gaussian operations.  

Here, we focus on two specific non-Gaussian operations: photon addition and subtraction \cite{ourjoumtsev_generating_2006, parigi_probing_2007,zavatta_experimental_2007,zavatta_subtracting_2008}. In the single-mode case, these processes are described and understood in a reasonably straightforward way (see e.g.~\cite{dakna_generating_1997}). Even though multimode scenarios prove to be much more challenging \cite{averchenko_multimode_2016}, mode-selective coherent photon subtraction is gradually coming within range \cite{ra_tomography_2017}. In two-mode setups these states have proven their potential, e.g., in the context of entanglement distillation \cite{ourjoumtsev_increasing_2007,kurochkin_distillation_2014,takahashi_entanglement_2010}. However the quantum properties of general multimode photon-added and -subtracted states remain unclear. 

In this letter, we present an exact and elegant expression for {\em Wigner functions} of the state obtained from the addition or subtraction of a single photon to a general multimode Gaussian state. We derive the conditions for achieving negativity in this Wigner function, which are needed for the states to potentially manifest a quantum advantage \cite{mari_positive_2012}. Moreover, we explain how the multiple modes in an experimental setup \cite{cai_reconfigurable_2016} can be entangled through mode-selective coherent photon addition or subtraction. For pure states, this entanglement is inherent in the sense that it cannot be destroyed by passive linear optics. \\

\paragraph{Optical phase space ---}The modal structure of light is essential throughout this work. In classical optics, a {\em mode} $u({\bf r}, t)$ is simply a normalised solution to Maxwell's equations. Multimode light is thus a sum of electric fields with complex amplitudes, $\sum_j (x_j + i p_j) u_j({\bf r}, t)$, associated with a specific mode basis $\{u_j({\bf r}, t)\}$. For each mode in this decomposition the real and imaginary part of the electric field are, respectively, the {\em amplitude and phase quadratures}. Thus, light comprised of $m$ modes, is described by $2m$ quadratures which are represented by a vector $f = (x_1, \dots x_m, p_1, \dots, p_m)^t \in \mathbb{R}^{2m}$. 

The same light can be represented in different mode bases, which boils down to changing the basis of $\mathbb{R}^{2m}$. This implies that any {\em normalised vector} $f \in {\cal N}(\mathbb{R}^{2m})$ can be associated with a single mode \footnote{We introduce the notion ${\cal N}(\mathbb{R}^{2m}) = \{f \in \mathbb{R}^{2m} \mid \norm{f} = 1\}$ to emphasise vectors associated with modes.}. However, the fact that quadratures always come in pairs induces additional structure on our space. This is described by a matrix $J$ that connects phase to amplitude quadratures and induces a symplectic structure. For this matrix, we have that $J^2 = - \mathbb{1}$ and $(J\!f_1,J\!f_2) = (f_1,f_2)$, for all $f_1, f_2 \in {\cal N}(\mathbb{R}^{2m})$, where $(.,.)$ denotes the innerproduct in $\mathbb{R}^{2m}$. Because of this symplectic structure, we now refer to $\mathbb{R}^{2m}$ as the {\em optical phase space}. Furthermore, the space generated by $f \in {\cal N}(\mathbb{R}^{2m})$, and its symplectic partner $J\!f$, is itself a phase space associated with a single mode.

The optical phase space is a basic structure from classical optics which must be quantised to study problems in quantum optics. To do so, we associate a quadrature operator $Q(f)$ to every $f \in {\cal N}(\mathbb{R}^{2m})$. To be compatible with different mode bases,  $Q(x_1 f_1 + x_2 f_2) = x_1 Q(f_1) + x_2 Q(f_2)$, must hold for any $x_1, x_2 \in \mathbb{R}$ and $f_1, f_2 \in {\cal N}(\mathbb{R}^{2m})$ such that $x_1^2+x_2^2=1$. In addition, they also obey the canonical commutation relations \cite{petz_invitation_1990,verbeure_many-body_2011}:
\begin{equation}\label{eq:CCR}
[Q(f_1),Q(f_ 2)] = - 2i(f_1, J\!f_2),
\end{equation} 
which are scaled to set shot-noise to one. Moreover, these quadrature operators are narrowly connected to the creation and annihilation operators, $a^{\dag}(g) = \big(Q(g) - i Q(J\!g)\big)/2$ and $a(g) = \big(Q(g) + i Q(J\!g)\big)/2,$ respectively. Note that $g \in {\cal N}(\mathbb{R}^{2m})$ denotes the mode in which a photon will be added or subtracted. One directly sees that $a(J\!g) = i a(g)$, relating the action of photon creation or annihilation on different quadratures of a two-dimensional phase space to different phases.\\

\paragraph{Truncated correlations ---} We use the density matrix $\rho$ to represent the quantum state and deduce the statistics of quadrature measurements. This letter focuses on multimode Gaussian states $\rho_G$, with expectation values denoted by $\<.\>_G$, which are de-Gaussified through the mode-selective addition or subtraction of a photon. These procedures induce new states given by
\begin{equation}\label{eq:photonSubRho}
\rho_{+} = \frac{a^{\dag}(g)\rho_G a(g)}{\<\hat{n}(g)\>_G+1}, \quad \text{and } \quad \rho_{-} =  \frac{a(g)\rho_G a^{\dag}(g)}{\<\hat{n}(g)\>_G},
\end{equation}
for addition and subtraction, respectively. The latter process has already been implemented experimentally \cite{ra_tomography_2017} following the recipe of \cite{averchenko_multimode_2016}. In line with these experiments, we will first assume that $\<Q(f)\>_G = 0$, such that the initial Gaussian state is not displaced. The remainder of this letter will deal with the characterisation of these quantum states. Our initial tool to do so is the {\em truncated correlation function}, recursively defined as
\begin{align}
\<Q(f_1)\dots Q(f_{n})\>_T = &\tr\{\rho\, Q(f_1)\dots Q(f_{n}) \}\\ &- \sum_{P \in {\cal P}}\prod_{I \in P} \<Q(f_{I_1})\dots Q(f_{I_{r}})\>_T \nonumber
\end{align}
where we sum over the set ${\cal P}$ of all possible partitions $P$ of the set $\{1,...,n\}$. In short, the n-point truncated correlation subtracts all possible factorisations of the total correlation. Hence, the truncated correlation functions are a multimode generalisation of cumulants. These functions are the perfect tools to characterise {\em Gaussian states}, since they have the property that $\<Q(f_1)\dots Q(f_{n})\>_T = 0$ for all $n > 2$ and all $f_1, \dots f_n \in {\cal N}(\mathbb{R}^{2m})$. On the other hand, this implies that non-Gaussian states must have non-vanishing truncated correlations of higher orders.

Through the linearity of the expectation value, we first calculate that the two-point correlation of photon-added (``$+$'') and -subtracted (``$-$'') states is given by
\begin{equation}\label{eq:A}
\begin{split}
&\<Q(f_1)Q(f_2)\>^{\pm} = \<Q(f_1)Q(f_2)\>_G + (f_1, A^{\pm}_{g}f_2),
\end{split}
\end{equation}
where $\<Q(f_1)Q(f_2)\>_G = (f_1, V\!f_2) - i (f_1, J\!f_2)$, with $V$ the Gaussian state's covariance matrix. The imaginary part of $\<Q(f_1)Q(f_2)\>_G$ is directly inherited from (\ref{eq:CCR}), whereas the final term in (\ref{eq:A}) is a consequence of the photon-subtraction process. A straightforward calculation identifies
\begin{equation}\label{eq:AMat}\begin{split}A^{\pm}_{g}= 2\frac{(V \pm \mathbb{1})(P_{g} + P_{J\!g})(V \pm \mathbb{1})}{\tr\{ (V \pm \mathbb{1})(P_{g} + P_{J\!g})\}},\end{split}
\end{equation}
where $P_g$ is the projector on $g \in {\cal N}(\mathbb{R}^{2m})$, such that $P_{g} + P_{J\!g}$ projects on the two-dimensional phase space associated with mode $g$. However, the two-point correlations (\ref{eq:A}) do not offer direct insight in the non-Gaussian properties of the state. Measuring higher order truncated correlations immediately shows a more refined perspective. Indeed, after some combinatorics, we obtain \cite{WalschaersTech} that, for all $k >1$
\begin{align}\label{eq:TruncFinalMix}
&\<Q(f_1)\dots Q(f_{2k-1})\>^{\pm}_T = 0,\\
&\<Q(f_1)\dots Q(f_{2k})\>^{\pm}_T = (-1)^{k-1} (k-1)! \\
&\qquad\qquad\qquad\qquad\qquad\quad\times\sum_{P \in {\cal P}^{(2)}}\prod_{I \in P} (f_{I_1}, A^{\pm}_{g}f_{I_2}),\nonumber
\end{align}
where ${\cal P}^{(2)}$ is the set of all pair-partitions \footnote{A pair-partition divides the set $\{f_1, \dots, f_{2k}\}$ up in $k$ pairs.}.
The prevalence of these correlations is immediately the first profoundly non-Gaussian characteristic of these single-photon added and subtracted multimode states.\\ 

\paragraph{Wigner function ---}While the truncated correlations themselves may provide good signatures of non-Gaussianity, they do not directly allow us to extract quantum features such as negativity of the Wigner function. However, they are directly connected to the Wigner function via the {\em characteristic function} $\chi(\alpha) = \tr (e^{i Q(\alpha)}\rho_{\pm}),$ for any point $\alpha \in \mathbb{R}^{2m}$ in phase space \footnote{Due to the linear structure $Q(x_1 f_1 + x_2 f_2) = x_1 Q(f_1) + x_2 Q(f_2)$, for any $x_1, x_2 \in \mathbb{R}$ and $f_1, f_2 \in \mathbb{R}^{2m}$, $Q(f)$ can be defined for non-normalised $f \in \mathbb{R}^{2m}$.}. It can be shown \cite{verbeure_many-body_2011} that this function can be written in terms of the cumulants:
\begin{equation}\label{eq:chiCumulants}
\chi(\alpha) = \exp\left( \sum^{\infty}_{n=1}\frac{i^n}{n!} \<Q(\alpha)^n\>_T\right).
\end{equation}
We then combine (\ref{eq:chiCumulants}) with (\ref{eq:TruncFinalMix}) to obtain the Wigner function as the Fourier transform of $\chi$, which leads to a particularly elegant expression, and the {\em key result} of this letter (see \cite{WalschaersTech} for technical details):
\begin{align}
W^{\pm}(\beta) 
 &= \frac{1}{2} \Big[(\beta, V^{-1} A^{\pm}_{g} V^{-1} \beta) -  \tr(V^{-1}A^{\pm}_{g}) + 2 \Big]W_0(\beta) \label{eq:WignerFunction},
\end{align}
where $\beta \in \mathbb{R}^{2m}$ can be any point in the optical phase space. $W_0(\beta)=(2\pi)^{-m} (\det V)^{-1/2} \exp\left(-(\beta, V^{-1}\!\beta)/2 \right)$ is the initial Gaussian state's Wigner function.\\



\paragraph{Entanglement ---} With the Wigner function (\ref{eq:WignerFunction}), we have the ideal tool at hand to study the quantum properties of multimode photon-added and subtracted states. First, we use it to investigate their separability under passive linear optics transformations. We will refer to a state as {\em passively separable} whenever we can find a mode basis where the state is {\em fully} separable, i.e.~where the Wigner function can be written as 
\begin{equation}\label{eq:Factorisation}\begin{split}
&W(\beta) =\int\!{\rm d}\lambda\, p(\lambda) W_{\lambda}^{(1)}(\beta^{(1)}_x, \beta^{(1)}_p) \dots W_{\lambda}^{(m)}(\beta^{(m)}_x, \beta^{(m)}_p),\\
\end{split}
\end{equation} 
with $p(\lambda)$ a probability distribution and $\lambda$ a way of labelling states. The $\beta^{(j)}_q$ are the coordinates of the vector $\beta$ in the symplectic basis where the state is separable. If no such symplectic basis exists, the state can never be rendered separable by passive linear optics, and we refer to it as {\em inherently entangled}.

We approach this question, starting from the initial Gaussian state $\rho_G$, which generally is mixed, characterised by the covariance matrix $V$.  This implies \cite{cerf_gaussian_2007} natural decompositions of the form $V= V_s + V_c$, with $V_c$ and $V_s$ interpreted as covariance matrices: $V_s$ is associated with a pure squeezed vacuum $\rho_s$, to which we add classical Gaussian noise given by $V_c$. There are many possible choices for such $V_s$ and $V_c$, which all allow for a rewriting of the Gaussian state in the form 
\begin{equation}\label{eq:identity}
\rho_G = \int_{\mathbb{R}^{2m}}{\rm d}^{2m}\xi\, D(\xi)\rho_s D^{\dag}(\xi) \frac{\exp\left(-\frac{(\xi, V_c^{-1}\xi)}{2}\right)}{(2\pi)^m \sqrt{\det V_c}},
\end{equation} 
where $D(\xi) = \exp(i Q(J\!\xi)/2)$ is the displacement operator. When we insert (\ref{eq:identity}) in (\ref{eq:photonSubRho}), we can now rewrite the photon-added or -subtracted Gaussian mixed state as a statistical mixture of photon-added or -subtracted displaced Gaussian pure states. After a cumbersome calculation invoking the commutation relations between creation, annihilation, and displacement operators, we find the following convex decomposition of the Wigner function (\ref{eq:WignerFunction}):
\begin{equation}\label{eq:theisdecomp}
W^{\pm}(\beta) = \int_{\mathbb{R}^{2m}}{\rm d}^{2m}\xi \, W^{\pm}_{\xi}(\beta) p^{\pm}_c(\xi),
\end{equation}
where 
\begin{equation}\begin{split}
p^{\pm}_c(\xi) = &\frac{\tr\big( (V_s + \norm{\xi}^2P_{\xi} \pm \mathbb{1})(P_{g}+P_{J\!g}) \big) e^{-\frac{(\xi, V_c^{-1}\xi)}{2}}}{\tr\big( (V \pm \mathbb{1})(P_{g}+P_{J\!g}) \big)(2\pi)^m \sqrt{\det V_c}},
\end{split}
\end{equation}
is a classical probability distribution. Indeed, it is straightforwardly verified that it is positive and normalised. In addition, the Wigner function for a displaced photon-added (``$+$'') or subtracted state (``$-$'') is found to be equal to \footnote{(\ref{eq:WignerDisp}) holds for general covariance matrices, and is therefore the general Wigner function of photon addition and subtraction from a displaced Gaussian state.}:
\begin{align}\label{eq:WignerDisp}
W^{\pm}_{\xi}(\beta)=  &\frac{W_{s}(\beta - \xi)}{\tr\big( (V_s + \norm{\xi}^2P_{\xi} \pm \mathbb{1})(P_{g}+P_{J\!g}) \big)} \\ &\times\Bigg( \norm{(P_g+P_{J\!g})(\mathbb{1}\pm V_s^{-1})  (\beta-\xi)}^2\nonumber \\&\qquad +  2\big(\xi, (P_g+P_{J\!g})(\mathbb{1}\pm V_s^{-1}) (\beta-\xi) \big) \nonumber\\&\qquad+ \tr\big((P_g +P_{J\!g})(\norm{\xi}^2P_{\xi} - V_s^{-1} \mp \mathbb{1} )\big) \Bigg)\nonumber.
\end{align}
$W_s$ denotes the Wigner function of the squeezed vacuum state with covariance matrix $V_s$. Because $W^{\pm}_{\xi}(\beta)$ is the Wigner function for a pure state, passive separability follows from the existence of a mode basis where $W^{\pm}_{\xi}(\beta)$ is factorised. 

Since  $W^{\pm}_{\xi}(\beta)$ represents the initial Gaussian state multiplied by a polynomial, it can only be factorised in the basis where $W_s(\beta)$ is factorised. The polynomial is fully governed by the vector $(P_g+P_{J\!g})(\mathbb{1}\pm V_s^{-1})  (\beta-\xi)$ which is contained in the two dimensional phase space associated with the addition/subtraction mode. Hence, $W^{\pm}_{\xi}(\beta)$ factorises if and only if the photon is added or subtracted to one of the modes that factorises $W_s(\beta)$. In other words, when we consider a pure Gaussian state in the mode basis where it is separable, we can induce entanglement by subtracting (or adding) a photon in a superposition of these modes. Moreover, it is impossible to undo the induced entanglement by passive linear optics. This induced entanglement is thus of different nature than gaussian entanglement, and is potentially important for quantum information protocols.

Furthermore, because (\ref{eq:theisdecomp}) is valid for every possible choice of $V_s$, we obtain that {\em the state is passively separable whenever the subtraction or addition takes place in a mode which is part of a mode basis for which the initial Gaussian state is separable}. For mixed initial states, it is unclear that subtraction or addition in a mode which is {\em not} part of such a mode basis automatically leads to inherent entanglement because also convex decompositions which are not of the form (\ref{eq:theisdecomp}) must be considered. Note that alternative methods exist to assess the entanglement of general CV states \cite{PhysRevA.74.030302, PhysRevA.90.052321}. However, these methods are not appropriate to gain analytical understanding of a whole class of states.


\begin{figure}
\centering
\includegraphics[width=0.4\textwidth]{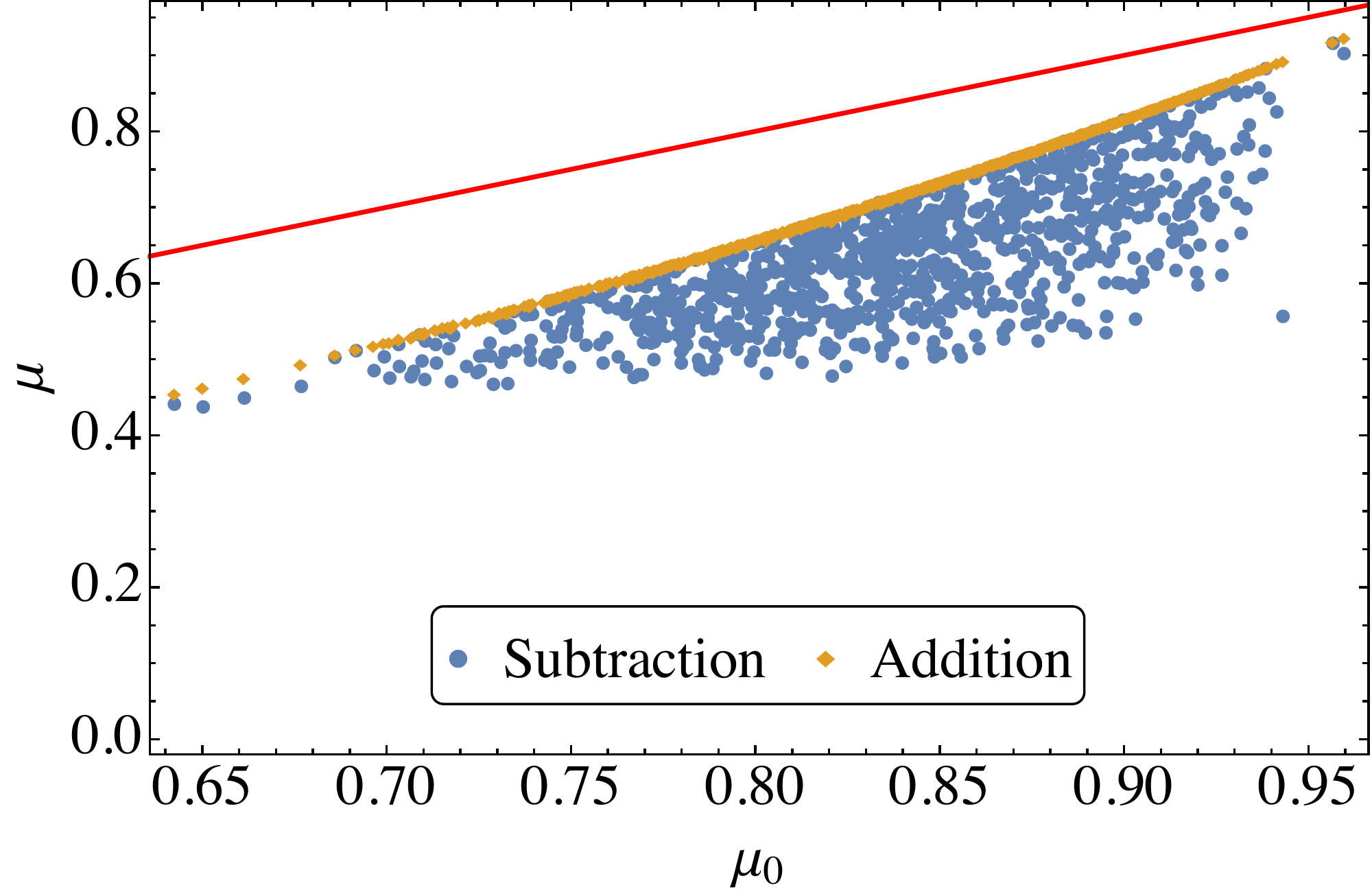}
\caption{Purities (\ref{eq:purity}) $\mu$ of Wigner functions for the reduced state, with all modes but mode $g$, in which addition or subtraction takes place, integrated out, compared to purities $\mu_0$ of the same mode's reduced state before photon addition or subtraction (i.e.~$\mu_0$ is obtained from the initial pure Gaussian state). Each point is a different realisation of a random choice for $g\in{\cal N}(\mathbb{R}^{2m})$, generated by choosing components from a standard normal distribution and subsequently normalising $g$. The red line indicates the cases where $\mu = \mu_0$. Lower purities imply higher entanglement\label{fig:Ent}}
\end{figure}

To illustrate the pure state result, we resort to an entanglement measure which is easily calculated from the Wigner function, the purity of a reduced state \cite{ozorio_de_almeida_entanglement_2009}. We study the entangling potential of photon subtraction and addition from a pure Gaussian state derived from an experimentally generated sixteen-mode covariance matrix $V_{\rm exp}$  \cite{cai_reconfigurable_2016}. We use the Williamson decomposition to separate $V_{\rm exp}$ into a pure multimode squeezed state $V^p_{\rm exp}$ and thermal noise, and ignore this thermal contribution \footnote{Thermal contribution $V_c$ is smaller than $V^p_{\rm exp}$ as quantified by the Hilbert-Schmidt norm: $\norm{V^p_{\rm exp}}_{\rm HS}/\norm{V_c}_{\rm HS} = 2.79773$.}. The squeezed mode basis of $V^p_{\rm exp}$ is referred to as the basis of {\em supermodes}. The single photon is added or subtracted in a random superposition of supermodes, characterised by a random $g \in {\cal N}(\mathbb{R}^{2m})$ 

In Fig.~\ref{fig:Ent}, we investigate the entanglement of mode $g$ to the rest of the system. We obtain the reduced state's Wigner function $W_{(g)}^{\pm}(\beta')$ (where $\beta'\in {\rm span}\{g, J\!g\}$) by integrating out all modes but the one associated with $g$. We then find the purity $\mu$ by evaluating \cite{ozorio_de_almeida_entanglement_2009} 
\begin{equation}\label{eq:purity}
\mu = 4\pi\int_{\mathbb{R}^2} {\rm d}^2\beta'\, \abs{W_{(g)}^{\pm}(\beta')}^2.
\end{equation}
The smaller the value $\mu,$ the stronger the mode $g$ is entangled to the remainder of the system. However, because we consider the entanglement of a superposition of supermodes to the remainder of the system, the mode $g$ will already be entangled in the initial Gaussian state. Therefore we also evaluate the purity $\mu_0$ obtained when the initial Gaussian state $W_0(\beta)$ is reduced to the the mode $g$. We see in Fig.~\ref{fig:Ent} that both addition and subtraction of a photon lower the purity of the reduced state, hence increasing the entanglement between the mode of subtraction/addition and the other fifteen modes, a multimode generalisation of what was observed for two modes \cite{kurochkin_distillation_2014}. Importantly, it is shown that photon subtraction typically leads to lower purities and thus distills more entanglement, which is in agreement with other recent work \cite{PhysRevA.93.052313}.\\


\paragraph{Wigner function negativity ---}Entanglement alone is, however, insufficient to reach a potential quantum advantage, we also require Wigner functions which are negative for certain regions of phase space  \cite{mari_positive_2012}. In pursuit of this goal, it is directly seen that the Wigner function (\ref{eq:WignerFunction}) becomes negative if (and only if) $(\beta, V^{-1} A^{\pm}_{g} V^{-1} \beta) -  \tr(V^{-1}A^{\pm}_{g}) + 2 < 0$ for some values of $\beta$. Because $(\beta, V^{-1} A^{\pm}_{g} V^{-1} \beta) \geqslant 0$, we can derive a particularly elegant {\em necessary and sufficient} condition for the existence of negative values of the Wigner function:
\begin{equation}\label{eq:character}
\begin{split}
&(g,V^{-1}g) + (J\!g,V^{-1} J\!g) > 2 \quad \text{for subtraction,}\\
&(g,V^{-1}g) + (J\!g,V^{-1} J\!g) > -2 \quad \text{for addition.}
\end{split}
\end{equation}
Through the combination of condition (\ref{eq:character}) with (\ref{eq:AMat}) we obtain a predictive tool that can be used to determine to (from) which modes $g \in {\cal N}(\mathbb{R}^{2m})$ a photon can be added (subtracted) to render the Wigner function negative. Note, moreover, that inequality (\ref{eq:character}) for photon addition always holds, implying that the Wigner functions of a single-photon added state is always negative.

\begin{figure}
\centering
\includegraphics[width=0.4\textwidth]{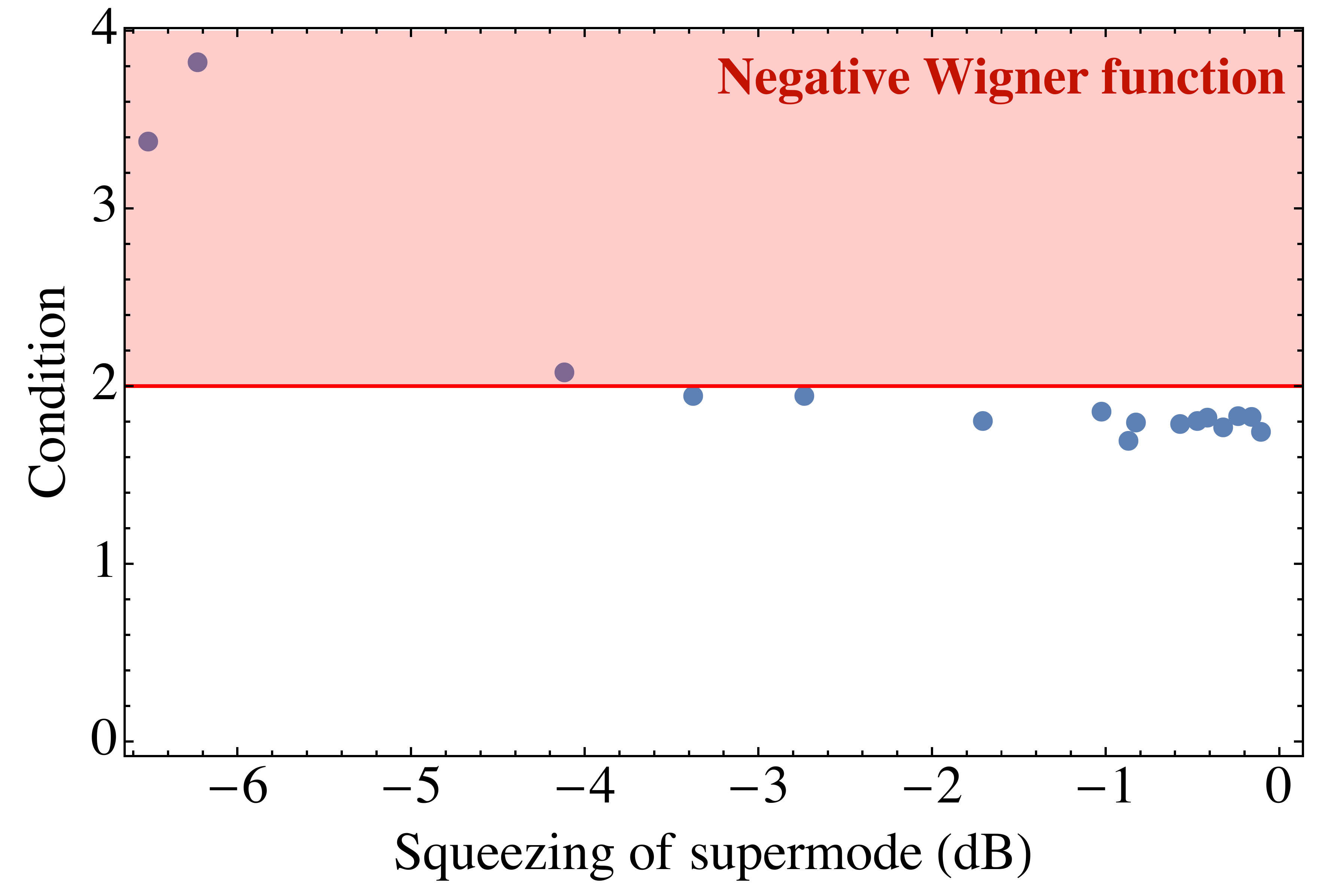}
\caption{Test of negativity condition (\ref{eq:character}) for an experimentally obtained \cite{cai_reconfigurable_2016} Gaussian state, with simulated photon subtraction in a supermode (points), as obtained through the Bloch-Messiah decomposition. For points falling in the red zone, photon subtraction in the associated supermode (see main text) leads to a negative Wigner function. The squeezing of the supermodes is indicated on the horizontal axis.\label{fig:NegSuper}}
\end{figure}

\begin{figure}
\centering
\includegraphics[width=0.4\textwidth]{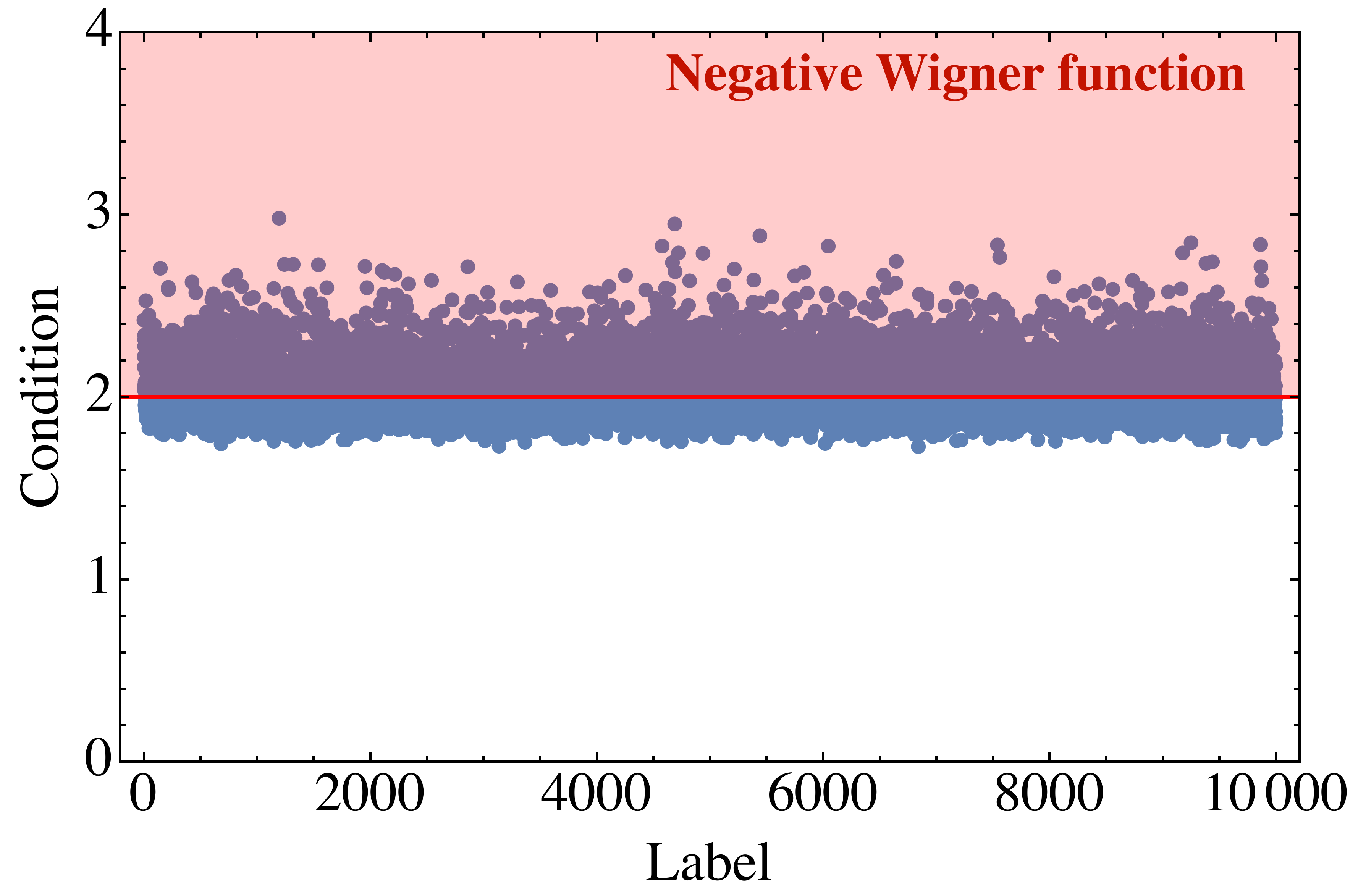}
\caption{Test of negativity condition (\ref{eq:character}) for an experimentally obtained \cite{cai_reconfigurable_2016} Gaussian state, with simulated photon subtraction in a random superposition of supermodes. Realisations falling in the red zone ($\approx 54\%$ of the realisations) have negative Wigner functions. \label{fig:NegRandom}}
\end{figure}

We can now study the condition (\ref{eq:character}) for the experimental state, characterised by $V_{\rm exp}$ in the case of photon subtraction, where the Wigner function is not guaranteed to be negative. In Fig.~\ref{fig:NegSuper} we subtract a single photon from a supermode, which only leads to negativity if the supermode is sufficiently squeezed (this is the case for merely three modes). Nevertheless, Fig.~\ref{fig:NegRandom} shows that subtraction from a coherent superposition of supermodes has an advantage regarding the state's negativity. For $54\%$ of the randomly chosen superpositions, i.e.~random choices of $g \in {\cal N}(\mathbb{R}^{2m})$, the Wigner function has a negative region. This underlines the potential of mode-selective photon subtraction to generate states with, both, a negative Wigner function, and inherent entanglement.\\

\paragraph{Conclusions --- } We obtained the Wigner function (\ref{eq:WignerFunction}) which results from mode-selective, pure addition or subtraction of a single photon to a non-displaced Gaussian state by exploiting truncated correlations (\ref{eq:TruncFinalMix}). We showed that subtraction and addition in a mode for which the initial Gaussian Wigner function takes the form (\ref{eq:Factorisation}), leaves the state {\em passively separable}, i.e.~any entanglement can be undone by passive linear optics. For a pure state, subtraction and addition of a photon in any other modes leads to {\em inherent entanglement}. It remains an open question whether this result can be generalised to mixed states. Moreover, we used the form (\ref{eq:WignerFunction}) to derive a practical witness (\ref{eq:character}) to predict whether the subtraction process induces negativity in the Wigner function (see also Figs.~\ref{fig:NegSuper} and \ref{fig:NegRandom}). Particularly relevant to current experimental progress is our conclusion that subtraction from a superposition of supermodes can produce inherently entangled states with non-positive Wigner functions, thus paving the road to quantum supremacy applications.

\begin{acknowledgements} \paragraph{Acknowledgements --- }This work is supported by the French National Research Agency projects COMB and SPOCQ, and the European Union Grant QCUMbER (no. 665148). C.F. and N.T. are members of the Institut Universitaire de France.
\end{acknowledgements}

\bibliography{Paper_NonGauss.bib}

\end{document}